# Optimization of reversible sequential circuits

## Abu Sadat Md. Sayem, Masashi Ueda

**Abstract**—In recent year's reversible logic has been considered as an important issue for designing low power digital circuits. It has voluminous applications in the present rising nanotechnology such as DNA computing, Quantum Computing, low power VLSI and quantum dot automata. In this paper we have proposed optimized design of reversible sequential circuits in terms of number of gates, delay and hardware complexity. We have designed the latches with a new reversible gate and reduced the required number of gates, garbage outputs, and delay and hardware complexity. As the number of gates and garbage outputs increase the complexity of reversible circuits, this design will significantly enhance the performance. We have proposed reversible D-latch and JK latch which are better than the existing designs available in literature.

———————————— ◆ ————————————

## 1 INTRODUCTION

Models of computation which are not logically reversible typically lose information in the process of execution. As the laws of physics appear to be reversible, that information cannot really be being lost, it must be being translated into another form. That form is usually heat. So, loss of information results power dissipation. To reduce this power dissipation reversible logic was introduced. The main idea of reversible logic is to allow the construction of reversible computers - by using components which preserve information content, and can thus potentially be run backwards. Hence, by implementing reversible designs of computer hardware significant amount of heat can be reduced. It has been shown that, for irreversible logic computations, each bit of information lost generates kTln2 joules of heat energy, where k is Boltzmann's constant and T the absolute temperature at which computation is performed [1],[2]. Benet showed the reverse that, kTln2 energy dissipation would not occur if the computation were carried out in a reversible manner [3].

Reversible circuits do not lose information and reversible computation in a system can be performed only when the system consists of reversible gates. Reversible logic is likely to be in demand in high speed power aware circuits, low-power CMOS design.

The main challenges of designing reversible circuits are to reduce number of gates, garbage outputs, delay and quantum cost. Another important matter is hardware complexity. In the existing designs in literature of sequential circuits several designs are proposed. In this paper we have proposed most optimized designs of reversible D-Latch and JK Latch. While designing the reversible latches; few researchers concentrated on reducing the number of gates and garbage output, while other tried to reduced the quantum cost. . In this paper we optimized the number of gates, garbage output, delay and hardware complexity for the total circuit and shown the results with illustrative calculation. Reversible RS and T latch is designed in the most optimized form in [4]. So we have worked with the D-Latch and JK Latch. A new reversible gate "Sayem Gate" (SG) is proposed here to design the latches.

## 2 REVERSIBLE LOGIC AND DIFFERENT REVERSIBLE GATES:

In this section, we have presented the basic definitions and ideas related to reversible logic and few reversible gates which are used and relevant with this research work.

**Definition2.1.a.** A *Reversible Gate* is a k-input, k-output (denoted by k * k) circuit that produces a unique output pattern [5], [6] for each possible input pattern.

**Definition 2.1.b.** *Reversible Gates* are circuits in which the number of outputs is equal to the number of inputs and there is a one to one correspondence between the vector of

• A. S. M. Sayem is with the National Institute of informatics, Tokyo, Japan.
• M. Ueda is with the National Institute of informatics, Tokyo, Japan.

inputs and outputs, i.e., it can generate unique output vector from each input vector and vice versa.

A reversible circuit must incorporate reversible gates in it and the number of gates used in a design is always a good complexity measure for the circuit. It is always desirable to realize a circuit with minimum number of gates.

**Example2.1.** Let the input vector be $I_v$, output vector $O_v$ and they are defined as follows, $I_v = (I_i, I_{i+1}, I_{i+2} \ldots I_{k-1}, I_k,)$ and $O_v = (O_i, O_{i+1}, O_{i+2} \ldots O_{k-1}, O_k)$. For each particular $i$, there exists the relationship $I_v \leftrightarrow O_v$

**Definition2.2.** Unwanted or unused output of a reversible gate (or circuit) is known as ***Garbage Output***. More formally, the outputs, which are needed only to maintain reversibility, are called *garbage outputs.*

**Example2.2.** If we wish to perform Exclusive-OR between two inputs, we can use the Feynman gate [7], but in that case, one extra output will be generated as well, which is the garbage output in this regard. The garbage output of Feynman is shown in Fig. 2.1 with *.

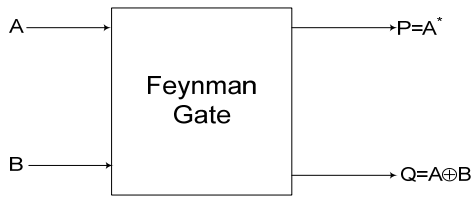

**Fig 2.1** Garbage Output

**Definition2.3.** The input vector, $I_v$ and output vector, $O_v$ for 2 * 2 *Feynman Gate (FG)* [7] is defined as follows: $I_v = (A, B)$ and $O_v = (P = A, Q = A \oplus B)$.

**Example 2.3:** The block diagram for 2 * 2 Feynman gate is shown in Fig 2.1.

Feynman gate is also known as CNOT (Controlled Not) gate. The two key reasons to use this gate in reversible circuit are:

i)      make the copy of an input ( putting any of the input a constant 0)
ii)     to invert an input bit ( putting any of the input a constant 1)

**Definition2.4.** The input vector, $I_v$ and output vector, $O_v$ for 3 * 3 *Toffoli gate (TG)* [8] is defined as follows: $I_v = (A, B, C)$ and $O_v = (P = A, Q = B, R = AB \oplus C)$.

**Example2.4.** The block diagram for 3 * 3 Toffoli gate is shown in Fig 2.2.

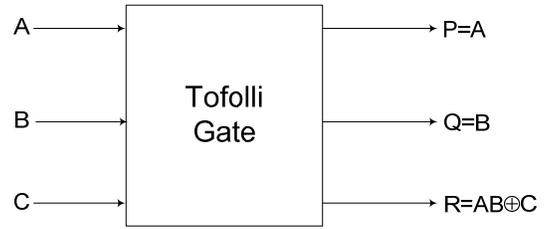

**Fig 2.2** Block diagram of Toffoli Gate

Toffoli gate plays an important role in the reversible logic synthesis. It is also used in the design of any Boolean function and hence it can be considered as a universal reversible gate.

**Definition2.5.** The input vector, $I_v$ and output vector, $O_v$ for 3 * 3 *Fredkin gate (FRG)* [9] is defined as follows: $I_v (A, B, C)$

= and $O_v = (P = A, Q = \bar{A} B \oplus AC, R = \bar{A} C \oplus AB)$.

**Example2.5.** The block diagram for 3 * 3 Fredkin gate is shown in Fig 2.3.

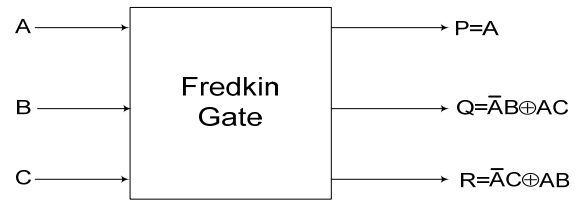

**Fig 2.3** Block diagram of a 3 * 3 Fredkin gate.

Fredkin gate also has its importance in reversible literature as it is a 1-through gate (one input is directly generated as output) and two other outputs can generate two different Boolean functions. Fredkin gate is the mostly used reversible gate to design reversible latches.

**Definition2.6.** The input vector, $I_v$ and output vector, $O_v$ for 3 * 3 *Peres gate (PG)* [10] is defined as follows: $I_v = (A, B, C)$ and $O_v = (P = A, Q = A \oplus B, R = AB \oplus C)$.

**Example2.6.** The block diagram for 3 * 3 Peres gate is shown in Fig 2.4

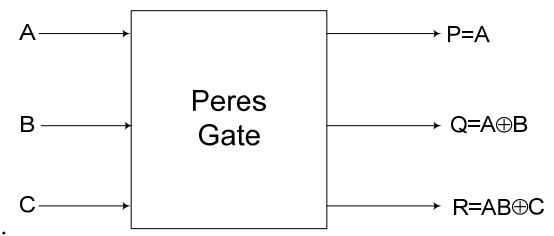

**Fig 2.4** Block diagram of a 3 * 3 Peres gate.

Actually the Peres Gate is the combination of Feynman Gate (FG) and Toffoli Gate (TG), and so it can simultaneously generate two output functions (from Q and R). Peres gate has the least quantum cost among the all 3X3 reversible gates, so it is mostly used in the various reversible logic circuits.

## 2.2 OPTIMIZATION PARAMETERS:

The main challenge of designing reversible circuits is to optimize the different parameters which result the design costly. The most important parameters which have dominant contribution in designing reversible circuits are

**Garbage Output:** Garbage outputs are the unwanted outputs of a reversible circuit which is described in the section 2.1.c. The less number of garbage outputs are produced the higher the performance and lesser the complexity of a circuit is.

**Number of gates:** The total number of gates used in a circuit. Minimum possible number of gates must be used in a circuit.

**Quantum Cost:** This refers to the cost of the circuit in terms of the cost of a primitive gate. It is calculated knowing the number of primitive reversible logic gates (1*1 or 2*2) required to realize the circuit.

**Flexibility:** This refers to the universality of a reversible logic gate in realizing more functions

**Gate Level:** This refers to the number of levels in the circuit which are required to realize the given logic functions.

**Hardware Complexity:** it refers to the total number of logic operation in a circuit. Means the total number of AND, OR and EXOR operation in a circuit. [11]

**Delay:** Delay is one of the most important parameter while designing reversible circuits. Many researchers suggested different definition of Delay for reversible circuits. In [4] Delay has been calculated by the quantum cost. In quantum computation each gate is realized using 1x1 or 2x2 reversible gates such Controlled -V gate,, Controlled-V+ gate, CNOT gate. In [4] if the quantum of cost of a reversible gate is x then the delay is considered as x$\Delta$, where $\Delta$ is unit delay .But as the delay is directly related to the number of gates so the concept of delay with quantum cost should yet to be taken under consideration. According to [12] delay is defined as follows

The delay of a logic circuit is the maximum number of gates in a path from any input line to any output line. This definition is based on the following assumptions:

- Each gate performs computation in one unit time.
- All inputs to the circuit are available before the computation begins.

The delay of the circuit of Fig 2.1 is obviously 1 as it is the only gate in any path from input to output.

In this paper in every calculation we use the definition of delay of [12].

## 3 PROPOSED REVERSIBLE GATE

We have proposed a new reversible gate named Sayem Gate (SG). SG is a 1 trough 4x4 reversible gate. The input and output vector of this gate are,

$Iv$= (A, B, C, D) and

$Ov$= (A, A'B $\oplus$ AC, A'B $\oplus$ AC $\oplus$ D, AB $\oplus$ A'C $\oplus$ D).the block diagram of this gate is shown in Fig 3.1

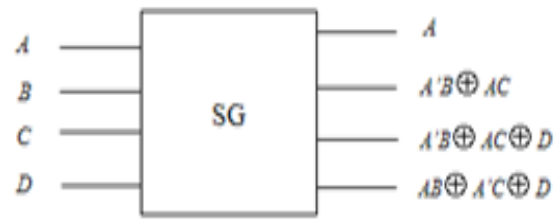

**Fig3.1**. Block diagram of new reversible SG

We can verify from the corresponding truth table of SG that the output and input vectors have one to one mapping between them which satisfies the condition of reversibility of a gate. The corresponding truth table is shown in Table I. We can see from Table I that the 16 different input and output vectors are unique means they have one to one mapping between them. So SG satisfies the condition of reversibility.

**TABLE I: TRUTH TABLE OF NEW REVERSIBLE SG**

| A | B | C | D | A | A'BΦ AC | A'BΦAC ΦD | ABΦA'C ΦD |
|---|---|---|---|---|---------|-----------|-----------|
| 0 | 0 | 0 | 0 | 0 | 0 | 0 | 0 |
| 0 | 0 | 0 | 1 | 0 | 0 | 1 | 1 |
| 0 | 0 | 1 | 0 | 0 | 0 | 0 | 1 |
| 0 | 0 | 1 | 1 | 0 | 0 | 1 | 0 |
| 0 | 1 | 0 | 0 | 0 | 1 | 1 | 0 |
| 0 | 1 | 0 | 1 | 0 | 1 | 0 | 1 |
| 0 | 1 | 1 | 0 | 0 | 1 | 1 | 1 |
| 0 | 1 | 1 | 1 | 0 | 1 | 0 | 0 |
| 1 | 0 | 0 | 0 | 1 | 0 | 0 | 0 |
| 1 | 0 | 0 | 1 | 1 | 0 | 1 | 1 |
| 1 | 0 | 1 | 0 | 1 | 1 | 1 | 0 |
| 1 | 0 | 1 | 1 | 1 | 1 | 0 | 1 |
| 1 | 1 | 0 | 0 | 1 | 0 | 0 | 1 |
| 1 | 1 | 0 | 1 | 1 | 0 | 1 | 0 |
| 1 | 1 | 1 | 0 | 1 | 1 | 1 | 1 |
| 1 | 1 | 1 | 1 | 1 | 1 | 0 | 0 |

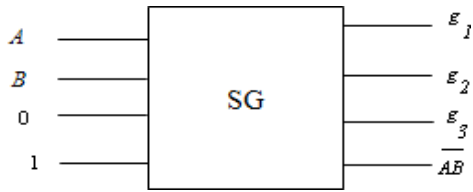

**Fig 3.2** SG as a two input universal gate

This gate can be used as a two input universal gate means it can perform any two input Boolean function. If we give the 3rd input 0 and 4th input 1 we get NAND of first two inputs at the 4th output which satisfies the universality of a gate in Boolean logic. This operation is shown in Fig 3.2. With this new reversible SG reversible Latches can be design efficiently.

## 4 DESIGN OF REVERSIBLE LATCHES:

In this section we described the proposed novel design of reversible D-Latch and JK Latch which are optimized in terms of the optimization parameters described in section 2.2.

### 4.1 D-LATCH:

The characteristic equation of D-Latch is $Q^+=DE+E'Q$. It can be realized with one SG. It can be mapped with SG by giving E, Q, D and 0 respectively in 1st, 2nd, 3rd and 4th input of SG. Fig 4.1(a) shows the design of D-Latch with only Q output and Fig 3.1(b) shows the design of reversible D-Latch with both the output Q and Q+ .One FG is needed to copy and produce the complement of Q from SG for the design of Fig 4.1(b). In the existing design of literature except the design of [4], [13], [14] all the latches were designed with only output Q. But the complement output Q+ is also needed in various logic implementation of nanotechnology based system. So we also proposed a novel design of reversible D-Latch with Q and Q+. Our design needs one SG and one FG while the design of [4] needs one FRG and two FGs. So numbers of gates are reduced. In [14] the same design is proposed using two FRG but in that case the hardware complexity is much than our design which is illustratively shown in the discussion section. The delay of this design is also less than the design of [4].

The gate required for the design with Q and Q+ is 2 while 3 gates are required in the design of

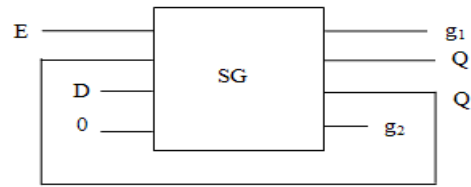

Fig 4.1(a): Proposed design of D-Latch with only output Q

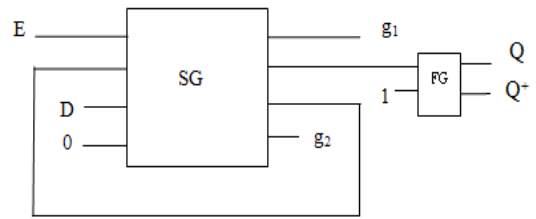

Fig 4.1(b): Proposed design of reversible D-Latch with output Q and Q+

[4]. Delay is calculated for this design with the calculation of [12] .Delay of our design is 2 while the delay of [4] is 3. This is the most optimized design of reversible D-Latch in terms of number of gates and garbage output. We observed that no further improvement is possible for number of gates required for the design of reversible D-Latch.

### 4.2 J-K LATCH:

The characteristics equation of JK latch is $Q^+=(JQ'+K'Q)E+E'Q$. It can be mapped with one FRG and 1 SG. Equation

$JQ'+K'Q$ is realized by 1 FRG with complement of K giving in the 3rd input. Now the 2nd output of FRG is $JQ'+K'Q$ which can be used as input D of D-latch shown in previous section. One SG is needed to get the desired output of $(JQ'+K'Q)E+E'Q$. For generating both the Q and Q+ we need another FG. Both designs are shown in Fig 4.2 (a) and Fig 4.2(b).

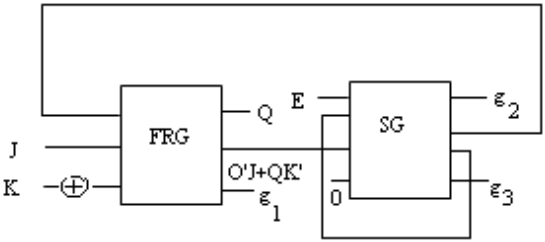

Fig 4.2(a) Proposed design of reversible JK Latch with output Q

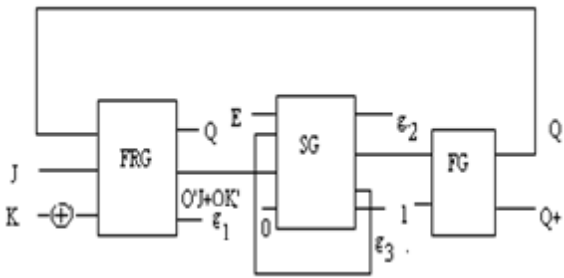

Fig 4.2(a) Proposed design of reversible JK Latch with output Q and Q+

To design the JK Latch with Only Q our design needs only 2 reversible gates with 3 garbage outputs and with both output Q and Q+ it needs 3 reversible gates with 3 garbage outputs.

# 5   RESULTS AND DISCUSSION

*5.1 Evaluation of Proposed Reversible SG:*

The proposed reversible gate is a 1 through 4X4 gate. It can be used as a two input universal gate. Using this gate the design of different Latches has been improved. As latches are most important memory elements and used in several circuits like RAM, Logic Blocks of FPGA [15] so this gate can contribute significantly in the reversible Logic community.

*5.2Evaluation of proposed design of D-Latch:*

The two different design of reversible D-Latch are optimized than the existing design of literature. In table II and III the comparison of different designs are shown.

**TABLE II: COMPARISON OF DIFFERENT D-LATCH WITH OUTPUT Q**

| D-Latch with Q | Components and cost | | |
|---|---|---|---|
| | *No of gates* | *Garbage Output* | *Delay* |
| This work | 1 | 2 | 1 |
| Existing work[4] | 2 | 2 | 2 |

**TABLE III: COMPARISON OF DIFFERENT D-LATCH WITH OUTPUT Q AND Q+**

| D-Latch with Q and Q+ | Components and cost | | |
|---|---|---|---|
| | *No of gates* | *Garbage Output* | *Delay* |
| This work | 2 | 2 | 2 |
| Existing work[4] | 3 | 2 | 3 |
| Existing work [13] | 7 | 6 | 7 |

From table II we can see that the design is better than the design of [4].here we need only one gate for generating the Q output while the design of [4] needs two gates. The delay of our design is 1 and the delay of [4] is 2. So this design is optimized than the design of [4] in terms of no of gates.

Table III shows the comparison of different D-latches with output Q and Q+. This work needs only 2 gates while [4] need 3 gates. The delay of our design is 2 and the delay of [4] is 3. So this design is better than [4] in terms of number of gates and delay. The same design is proposed with two gates in [14] but that design has more hardware complexity. If

$\alpha$ = A two input EX-OR gate calculation
$\beta$ = A two input AND gate calculation
$\delta$ = A NOT calculation

Then the hardware complexity of our design is $5\alpha+6\beta+3\delta$ where the hardware complexity of [14] is $4\alpha+8\beta+4\delta$ so this design is better than the design of [14] also.

*Evaluation of proposed design of Reversible JK Latch:*

Our proposed design of reversible JK latch is better than the design of existing designs in literature. Table IV and V shows the comparative results of different JK latch designs.

**TABLE IV: COMPARISON OF DIFFERENT REVERSIBLE JK LATCH WITH**

OUTPUT Q

| JK Latch with Q | Components and cost | | |
|---|---|---|---|
| | No of gates | Garbage Output | Delay |
| This work | 2 | 3 | 2 |
| Existing work[4] | 3 | 3 | 3 |

**TABLE V: COMPARISON OF DIFFERENT REVERSIBLE JK LATCH WITH OUTPUT Q AND $Q^+$**

| D-Latch with Q and Q+ | Components and cost | | |
|---|---|---|---|
| | No of gates | Garbage Output | Delay |
| This work | 3 | 3 | 3 |
| Existing design[4] | 4 | 3 | 4 |
| Existing design [13] | 10 | 12 | 10 |

From table IV we can see that the design of reversible JK latch with only output Q was realized by 3 gates and delay was 3 in [4] while our design is realized with two reversible gates and delay is 2. So this design is better than the design of [4] in terms of number of gate and delay.

The design of reversible JK latch with output Q and Q+ was realized with 4 gates in [4] and 10 gates in [13] where our design needs only 3 gates. The delay for our proposed design is 3, where in [4] its 4 and in [13] its 10. So our proposed design is optimized than [4] and [13].

The same design of reversible JK Latch was proposed with 3 gates in [14] but that design has much hardware complexity than ours. The hardware complexity of our design is $7\alpha+10\beta+7\delta$ and the hardware complexity of [14] is $6\alpha+12\beta+8\delta$. Though $\alpha$ is greater but the other two parameter are less than [14] thus we can say that our proposed design is better than [14] in terms of hardware complexity. We have observed that no further improvement is possible for designing a reversible JK latch in terms of number of gates.

## CONCLUSION:

In this paper we have proposed the reversible design of D-Latch and JK Latch. Latches are important memory element. Thus this optimization will result in great contribution in designing logic circuits with memory and sequential elements. We have given the lower bounds for both the design in terms of number of gates and delay. We have proposed a new reversible gate which can contribute significantly in reversible logic community.

## ACKNOWLEDGEMENT

Authors would like to thank National Institute of Informatics, Tokyo, Japan for supporting the research program. We are grateful to member of informatics lab of National Institute of Informatics for their co operation to complete the research work.

## REFERENCES

[1]  R. Keyes, R. Landauer "Minimal Energy Dissipation "In IBM Journal of Research and Development 1970; 14: 153-7.

[2]  R. Landauer Irreversibility and heat generation in the computational process's IBM Journal of Research Development 1961; 5: 183-91.

[3]  C.H Bennett "Logical reversibility of computation"; IBM Journal of Research and Development 1973; 17: 525-32.

[4]  R. Thapliyal, N. Ranganathan. Design of Reversible Latches Optimized for Quantum Cost, Delay and Garbage Outputs, vlsid, pp.235-240, 2010 23rd International Conference on VLSI Design, 2010

[5]  .M.H. Babu, M.R. Islam, A.R. Chowdhury, S.M.A. Chowdhury, "Reversible logic synthesis for minimization of full-adder circuit", IEEE Conference on Digital System Design 2003; 50-4.

[6]  H.M.H. Babu, M.R. Islam, A.R. Chowdhury, .SM.A. Chowdhury, "Synthesis of full-adder circuit using reversible logic", 17th International Conference on VLSI Design 2004; 757-60.

[7]  R. Feynman, Quantum Mechanical Computers, Optical News 1985; 11-20.

[8]  T.Toffoli "Reversible Computing", Tech memo MIT/LCS/TM-151, MIT Lab for Computer Science 1980.

[9]  E. Fredkin and E. Toffoli "Conservative Logic", International Journal of Theoretical Physics, 1983; 21: 219-53.

[10] A. Peres. "Reversible Logic and Quantum Computers", Physical Review, 1985; 3266-76.

[11] M. Haghparast,K. Navi," A Novel Reversible BCD Adder For Nanotechnology Based Systems"; American Journal of Applied Sciences 5 (3): 282-288, 2008,ISSN 1546-9239.

[12] A. K. Biswas, M. M. Hasan, A.R. Chowdhury and H. M. H. Babu, Efficient Algorithms for Implementing Reversible Binary Coded Decimal Adders, *Microelectron. J*, 39(12):1693–1703, 2008.

[13] H. Thapliyal, M. B. Srinivas, and M. Zwolinski," A beginning in the reversible logic synthesis of sequential circuits",In *Proc. the Military and Aerospace Programmable Logic Devices Intl. Conf.*, Washington, Sept. 2005.

[14] H. Thapliyal and A. P. Vinod, "Design of reversible sequential elements with feasibility of transistor implementation" In *Proc. the 2007 IEEE Intl. Symp. On Cir. and Sys.*, pages 625–628, New Orleans, USA, May 2007.

[15] A.S.M. Sayem,M.M.A. Polash,H.M.H. Babu,"Design of a reversible logic block of FPGA", proceedings of silver Jubilee Conference on Communication Technologies and VLSI design (CommV'09), VIT University, Vellore, India.Oct. 8-10, 2009, pp: 501-502.

A.S.M. Sayem received his B.Sc. degree in Computer Science & Engineering from University of Dhaka, Bangladesh in 2009.He is perusing his M.Sc.in Computer Science & Engineering from University of Dhaka and currently on a research program in National Institute of Informatics, Tokyo, Japan. He is a member of IEEE. He was awarded IEEE Computer Society best student paper award in the " Silver Jubilee Conference on Communication Technologies and VLSI design (CommV'09), VIT University, Vellore, India.Oct. 8-10, 2009." His research interest includes reversible logic design, Low power VLSI, DNA computing and finite languages.

M.Ueda is an Assistant Professor in National Institute of Informatics, Japan from 2006.  He received his B.Ec. in Faculty of Economics, Kyoto University in 1998, and M.I. in Graduate School of Informatics, Kyoto University in 2000.  He was a visiting fellow in Crawford School of Economics and Government, Australian National University in 2006, and was post doctorial fellow in Institute of Economics and Political Studies, Kansai University, Japan from 2003 to 2005.  His concern is social effect of network structure and its competitive modeling.  His analysis is focused upon public utility services, like high-speed Internet infrastructure, grid networks, and open source software, based upon economics and management approach.